\documentclass[8.5pt,twoside,twocolumn]{article}
\usepackage[super,sort&compress,comma]{natbib} 
\usepackage{times,mathptmx}
\usepackage{sectsty}
\usepackage{balance} 

\usepackage{graphicx} 
\usepackage{lastpage}
\usepackage[format=plain,justification=raggedright,singlelinecheck=false,font=small,labelfont=bf,labelsep=space]{caption} 
\usepackage{fancyhdr}
\begin{document}

\thispagestyle{plain}
\fancypagestyle{plain}{
\renewcommand{\headrulewidth}{1pt}}
\renewcommand{\thefootnote}{\fnsymbol{footnote}}
\renewcommand\footnoterule{\vspace*{1pt}%
\hrule width 3.4in height 0.4pt \vspace*{5pt}} 
\setcounter{secnumdepth}{5}

\makeatletter 
\def\subsubsection{\@startsection{subsubsection}{3}{10pt}{-1.25ex plus -1ex minus -.1ex}{0ex plus 0ex}{\normalsize\bf}} 
\def\paragraph{\@startsection{paragraph}{4}{10pt}{-1.25ex plus -1ex minus -.1ex}{0ex plus 0ex}{\normalsize\textit}} 
\renewcommand\@biblabel[1]{#1}            
\renewcommand\@makefntext[1]%
{\noindent\makebox[0pt][r]{\@thefnmark\,}#1}
\makeatother 
\renewcommand{\figurename}{\small{Fig.}~}
\sectionfont{\large}
\subsectionfont{\normalsize} 

\renewcommand{\headrulewidth}{1pt} 
\renewcommand{\footrulewidth}{1pt}
\setlength{\arrayrulewidth}{1pt}
\setlength{\columnsep}{6.5mm}
\setlength\bibsep{1pt}

\makeatletter
\DeclareRobustCommand\onlinecite{\@onlinecite}
\def\@onlinecite#1{\begingroup\let\@cite\NAT@citenum\citealp{#1}\endgroup}
\makeatother 

\twocolumn[
  \begin{@twocolumnfalse}
\noindent\LARGE{\textbf{Materials Design by Quantum-Chemical and other Theoretical/Computational Means:
Applications to Energy Storage and Photoemissive Materials}}
\vspace{0.6cm}

\noindent\large{\textbf{K\'aroly N\'emeth$^{\ast}$\textit{$^{a}$} }}
\vspace{0.5cm}


\noindent \textbf{\small{DOI: 10.1039/b000000x}}
\vspace{0.6cm}

\noindent \normalsize{
The present paper discusses some recent developments in the field of rational design for energy
storage and photoemissive materials. Recent and new examples of designer materials for Li-ion and Li-air 
type batteries with high capacity and energy/power density as
well as photoemissive materials with low workfunctions and improved brightness are discussed as illustrative
examples of how quantum-chemical and other theoretical computational means can be used for rational materials
design.
}
\vspace{0.5cm}
 \end{@twocolumnfalse}
  ]

\section{Introduction}
Rational materials design has been one of the holy grails of chemistry and physics. With the advent of efficient
algorithms that solve the Schr\"odinger equation at various levels of approximation in (near) linear scaling
computational effort \cite{LinearScaling} as compared to the system size,
combined with parallel computers using ever faster processors, this goal is now within reach and the fruits of
decades of hard work in the development of such methodologies are increasingly harvested in terms of rationally
designed materials. Major projects funded by governments or by the industry invest substantial funds in rational 
materials design. The US government funds the Materials Genome Initiative to computationally explore, 
design and experimentally test designer materials with approximately 100 million dollars 
through
various agencies, such as NSF and DOE. Major materials design efforts are also underway in other programs as well,
for example within the Joint Center of Energy Storage Research (``Battery Hub'') program 
of DOE as electrode materials and electrolyte design efforts.
Similar initiatives can be seen in the private industry as well. For example,
Bosch, a major supplier of electrical appliances conducts systematic computational 
materials screening and design efforts in the fields of energy storage and conversion. 
Major companies, such as IBM, Bosch, and Apple
produce thousands of patent applications each year. More than half a million patent
applications are filed in the US annually, and about a quarter of a million patents are granted.  
There is an increasing number of patents solely based on theoretical materials design 
(``conceptual reduction to practice'') without experimental testing (``actual reduction to practice'') 
that can also be attributed to the increased competition in the race for
materials that can potentially control important emerging fields of applications, such as
energy storage, clean energy production, etc. 
\cite{BoschCO2patent10,CFx-ContourEnergy11,US8389178}. Patents have always been formulated so that they cover an as broad as
possible range of potential modifications of the core invention, leading sometimes to extremely
broad patents, such as Intel's one on all nanostructures smaller than 50 nm in diameter or Rice University's one
on all objects composed at least 99\% of carbon nanotubes \cite{JMPearce12}.

The vast majority of the materials design efforts for the exploration of crystalline materials use 
electronic structure codes with periodic boundary conditions 
such as VASP \cite{VASP} or Quantum Espresso \cite{QE}. These codes utilize plane wave
representation of wavefunctions, effective pseudoptentials to model
core electrons and various exchange and
correlation functionals within the DFT approach. While these codes and the underlying 
methods have sever limitations
for certain phenomena, such as electron correlation effects, band gaps in transition metal compounds,
intermolecular interactions \cite{JPaier06}, excited state properties, etc., 
effective core potentials, exchange/correlation potentials and novel DFT correction methodologies have been
developed to cure some of these problems.
These latter methods include the DFT+U \cite{dftu} method 
that turned out particularly practical for studying band gaps, thermo/electro-chemistry, 
phase changes, magnetic states and structural properties of solids.
Gaussian and mixed Gaussian-plane-wave representation based codes, as well as
wavelet or adaptive grid based ones
are also used in the exploration of materials. Such codes include
SIESTA \cite{Siesta}, Gaussian \cite{Gaussian09}, PQS \cite{PQS}, FreeON \cite{FreeON}, to name a few.
Some of these codes are also capable to calculate explicit electron-electron correlation with periodic
boundary conditions, for example utilizing quasiparticle (electron-pair) bands on the MP2 level of correlation
and beyond \cite{SSuhai83}.

Other approaches to materials design may be based on the ``mining'' and analysis of existing data deposited in
the scientific literature, with the aim of revealing potentially useful but yet unobserved or unutilized
connections and thereby constructing new structures/compositions of materials that are potentially superior
to existing and well observed ones \cite{KNemeth13}.

The present perspectives paper will focus on some recent developments in the field of materials design for 
Li-ion and Li-air batteries first, then describes approaches about the development of improved photoemissive
materials.

\section{Results and Discussion}   
\subsection{Li-ion batteries}
Most Li-ion batteries operate through the intercalation and deintercalation of Li-ions in the positive electrode
(cathode of the discharge process) electroactive materials. There is a great effort to optimize the properties of
the electroactive crystals. The optimization parameters include the gravimetric and volumetric energy densities
of the electroactive materials, their capacities (concentration of Li that can intercalate), the power densities
(how fast the charging and discharging may occur), the voltage associated with the corresponding electrochemical reaction,
the electrolytes that have to sustain the voltages and currents that occur and the economy of the materials involved, to
name a few optimization parameters. Most design efforts focus on layered crystalline cathode materials, such as spinels or
polyanionic compounds \cite{BCMelot13,GHautier11}. 
In order to computationally predict the voltage associated with an electrochemical cell reaction, 
one has to calculate the Gibbs free energy change associated with the reaction, express it in terms of eV-s and divide
by the number of electrons transferred during the reaction per molecule of product. The reaction energy is almost always
approximated as the difference of the electronic energies of the products and reactants which is usually a good
approximation when no gas molecules are involved that would cause large entropy changes \cite{FZhou04,AJain11,GHautier11}. 
However, the accurate
calculation of reaction energies and other properties of solids is often problematic, especially in case of transition
metal compounds.
It turns out, that DFT(GGA)+U methods are capable to provide reaction energies and voltages that compare well
with experiments for this class of materials \cite{FZhou04,AJain11,GHautier11}. 
Perhaps the best intercalation material designed for cathode applications so far, 
is Li$_{3}$Cr(BO$_{3}$)(PO$_{4}$) a polyanionic
material with sidorenkite (Na$_{3}$MnPO$_{4}$CO$_{3}$) structure \cite{GHautier11}. This material has not been synthesized yet. 
Lithium intercalation in the deintercalated Cr(BO$_{3}$)(PO$_{4}$) structure occurs at calculated voltages of 4.2-5.1 V,
relative to Li/Li$^{+}$, resulting in theoretical energy densities of 1705 Wh/kg and 4814 Wh/L with capacities of
354 mAh/g \cite{GHautier11}. 
For comparison, LiCoO$_{2}$ cathode materials in current Li-ion batteries
have a theoretical energy density of 568 Wh/kg (2868 Wh/L) and charge capacity
of 273 mAh/g (1379 mhA/cm$^{-3}$) \cite{US8389178}.

While the energy density of Li$_{3}$Cr(BO$_{3}$)(PO$_{4}$) and related materials is very attractive, 
their power density (the rate of charging and discharging) remains low, as the process of intercalation/deintercalation
is relatively slow, rendering the charging of Li-ion batteries typically an overnight long process. To develop
materials with both high power and energy densities,
the present author has recently proposed \cite{BNpatent} the use of functionalized hexagonal 
boron nitride (h-BN) monolayers as intercalation materials.
In this case the intercalation would happen into  
the surface of a 2D material directly from the electrolyte, avoiding the
slow diffusion inside the crystallites. 
The charging process would be similarly fast, allowing for large current densities. 
Here, two materials based on BN are briefly discussed. 
The first one is a 3D material, Li$_{3}$BN$_{2}$, the second one is a 2D one,
BNCN, a cyano (-CN) group functionalized h-BN.

\subsubsection{Li$_{3}$BN$_{2}$}
\begin{figure}[tb!]
\resizebox*{3.4in}{!}{\includegraphics{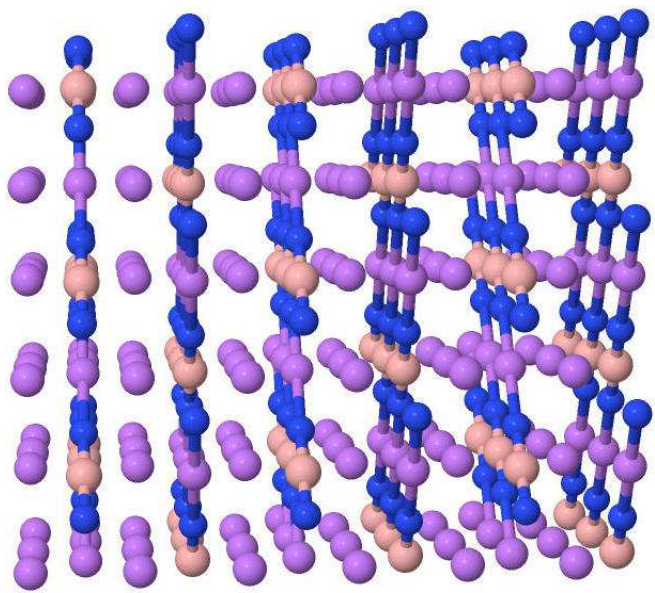}}
\caption{
Perspective view of the layered structure of $\alpha$-Li$_{3}$BN$_{2}$ in a 3x3x3 supercell. 
Color code: Li - violet, B - magenta, N - blue.
}
\label{Li3BN2}
\end{figure}

It has been known for decades
that the reaction of molten Li$_{3}$N with h-BN can produce various phases of the Li$_{3}$BN$_{2}$ 
crystalline material \cite{HYamane87}. $\alpha$-Li$_{3}$BN$_{2}$ has a
layered structure as shown in Fig. \ref{Li3BN2}, 
with two Li ions being mobile per formula unit, while the third Li is part
of a 1D rod-like polymer, with repeating -Li-N-B-N- sequence. The polymeric chains and the mobile Li ions are placed in
separate layers in the $\alpha$-Li$_{3}$BN$_{2}$ phase (space group P4$_{2}$/mnm). The cell reaction proposed
\cite{BNpatent} is 
2Li + LiBN$_{2}$ $\rightarrow$ Li$_{3}$BN$_{2}$ on discharge. 
The formal charge of the BN$_{2}^{n-}$ linear anion changes from n=-1 to n=-3 on discharge.
Since no heavy transition metal atoms are involved, one may
use the PBEsol functional \cite{PBE,PBEsol} to obtain realistic optimum crystal structures and reaction energies.   
The Quantum Espresso \cite{QE} software has been used, with ultrasoft pseudopotentials and 50 rydberg wavefunction cutoff,
using a 6x6x6 k-space grid.
The residual forces were smaller than 1.d-4 Ry/bohr at the optimum structures. Experimental 
lattice parameters a=b and c have been
reproduced with an accuracy of smaller than 1\% and 2.5\% errors, respectively. Note that the c-direction is perpendicular to the
layers mentioned. In the Li-deintercalated structure, the lattice parameters a=b and c become 1.2 and 3.3 \% shorter, which
indicates a cell volume shrinking of 5.6 \%, or 2.8\% per two-electron transfer. 
Note that such a relatively small cell volume change counts as acceptable for Li-ion battery electrode applications
\cite{GHautier11}. The cell reaction energy is calculated to be ${\Delta}E$ = -7.23  eV, indicating a cell voltage of 
U=3.61 V (assuming an Li/Li$^{+}$ anode). The gravimetric and volumetric energy densities are 3247 Wh/kg and 5919 Wh/L, respectively. These values are
significantly larger than those obtained for Li$_{3}$Cr(BO$_{3}$)(PO$_{4}$) \cite{GHautier11}. Especially the gravimetric energy density is almost
twice as large for Li$_{3}$BN$_{2}$ as for Li$_{3}$Cr(BO$_{3}$)(PO$_{4}$). Li$_{3}$BN$_{2}$ appears to be greatly superior to
Li$_{3}$Cr(BO$_{3}$)(PO$_{4}$) also in terms of gravimetric and volumetric capacity densities with the respective values of
899 mAh/g and 1638 mAh/cm$^{3}$. 
It is surprising that Li$_{3}$BN$_{2}$ has not been tested as cathode electroactive material yet. The reason is
probably that only the alpha phase can be expected to preserve its layered structure after the deintercalation of
mobile Li-ions. The monoclinic phase has been considered recently 
for application as part of a conversion based anode material \cite{THMason11}.
The high melting point of $\alpha$-Li$_{3}$BN$_{2}$, about 900 $^{o}$C \cite{HYamane87} 
is also indicative of the stability of the polymeric chains with -Li-N-B-N- repeating units, as the Li ions that
are not incorporated in the chains are known to be very mobile as excellent Li-ion conductivity values indicate
\cite{HYamane87}. Therefore, the present calculations also predict the existence of stable phases with 
the stoichiometry of Li$_{x}$BN$_{2}$ with $1<x<3$. Furthermore, potentially also the beta and the monoclinic phase would
be transformed to $\alpha$-Li$_{3}$BN$_{2}$ after repeated charge/discharge cycles, as the alpha phase
appears as the energetically most favorable packing of the polymeric -Li-N-B-N- rods.

\subsubsection{BNCNNa}
\begin{figure}[tb!]
\resizebox*{3.4in}{!}{\includegraphics{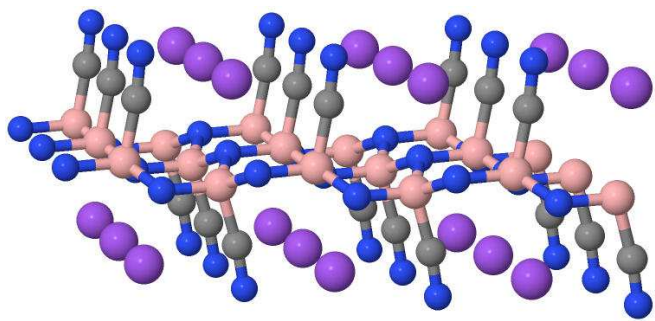}}
\caption{
Perspective view of the cyano functionalized h-BN monolayer with intercalated Na ions in a 3x3x1 supercell. 
Color code: Na - violet, B - magenta, N - blue, C - gray.
}
\label{BNCNNa}
\end{figure}

The reaction of molten sodium cyanide with h-BN is expected to lead to the cyano-functionalization of B sites on
both sides of the h-BN monolayers which then
intercalate Na$^{+}$ ions, as depicted in Fig. \ref{BNCNNa}. This structure is expected to lead to
fast charging and discharging due to the sterically unhindered access of Na$^{+}$ ions to the intercalation sites.
Using the above methodology and a $\approx$ 30 {\AA} vacuum layer to separate layers of BNCNNa, the electrochemical
properties of BNCNNa have been computed. Both intercalated and deintercalated structures are stable and preserve
the covalent cyano functionalized B centers. With Li atoms instead of Na,
the covalent functionalization would break up and LiCN and h-BN would form. The voltage of the Na + BNCN
$\rightarrow$ BNCNNa electrochemical cell is calculated to be U = 2.87 V, associated with a cell reaction energy
of 2.87 eV (assuming an Na/Na$^{+}$ anode). 
The gravimetric and volumetric energy densities are 1042 Wh/kg and 3547 Wh/L, 
the capacity values 363 mAh/g and 1236 mAh/cm$^3$. These values indicate great potential for example for use in
portable electronics devices. BNCN and analogous functionalized h-BN compunds can serve as universal intercalation
electrode materials capable to intercalate alkali, alkaline earth, Al and other cations (unless conversion reactions
happen) thus allowing for batteries based on the transfer of cations other than Li$^{+}$.

Note that in principle similar functionalization of graphene can also be envisioned, however the patterned and
polarized h-BN surface is a much better candidate for electrophile and nucleophile attack by functionalization
agents, such as molten salts. In case of graphene, all atoms are equivalent for functionalization, while for h-BN
only half of the atoms can be subject of, say, nucleophile attack, leaving the other half unfunctionalized providing
space for intercalation of cations and free N atoms in the BN surface to contribute to the complexation 
and binding of the intercalating ions as well as to the storage of the extra negative charge per formula unit 
due to discharge. 
Therefore, h-BN appears to be a better candidate to build 2D intercalation
structures than graphene. Doped graphene structures may have similar properties to h-BN for covalent
functionalization. Functional groups should be selected based on whether they make h-BN a strong electron acceptor
after the functionalization, when positive electrode materials are designed. For negative electrode 
materials, the functional group should be selected such that the resulting layer will be a good electron donor.
These considerations are analogous to the design of charge transfer salts \cite{JHuang08}.

While transition metal compounds currently dominate Li-ion battery electroactive materials, there is a lot
of potential in conjugated $\pi$-electron systems and their functionalized derivatives to be utilized as
electroactive materials. Concepts of these systems, such as the principles of charge transfer compounds, 
have practically been completely ignored in energy storage development so far. As a direct continuation of the
present work, a great variety of functionalization of h-BN will be explored to find the best novel anode and
cathode materials of this class of compounds. Band structures, conductivities, charge
distributions and structural changes upon intercalation/deintercalation will be analyzed to understand the
mechanism of charge storage in these materials. The experimental testing of some of
these materials is already underway through collaborative work at 
Illinois Institute of Technology and elsewhere.

\subsection{Li-air batteries}
Li-air batteries are considered as the ultimate large energy density batteries that would enable long range all
electric vehicles. They produce electrical energy through the reaction of metallic 
Li with O$_{2}$ taken from the air, whereby the discharge product must be Li$_{2}$O$_{2}$ in order the
battery be rechargeable. The fact that Li$_{2}$O$_{2}$ is an aggressive oxidant and may explosively oxidize the
carbon electrode in which it forms and deposits makes Li-O$_{2}$/peroxide batteries unsafe. A thorough analysis
\cite{KNemeth13} by the present author
about the thermochemistry and kinetics of Li-oxalate, Li$_{2}$C$_{2}$O$_{4}$, 
with the availability of catalysts for selective
and energy-efficient CO$_{2}$/oxalate conversions indicates that 
instead of O$_{2}$, CO$_{2}$ should be taken from air (or from a tank)
to construct a high energy and power density battery that can be safely used and is of environmentally benign and
economic materials. The details of the design based on existing experimental 
data can be found in Ref. \onlinecite{KNemeth13}.
A key component of the energy-efficiency of CO$_{2}$/oxalate cathodes is the catalyst that reduces the
overpotential of CO$_{2}$ reduction to nearly zero. While such catalysts are known, for example copper
complexes or oxygen molecule in selected ionic liquid electrolytes \cite{KNemeth13}, 
the applicability of these catalysts may vary in
the various implementations, therefore there is a need for the development of additional robust
catalysts to optimize the performance of metal-CO$_{2}$/oxalate batteries.

\subsection{Photoemissive materials}
Another example of materials design discussed here concerns photoemissive materials. Improved photocathodes
are needed for future electron and light sources, such as x-ray free electron lasers, energy recovery
linacs and dynamic and ultrafast transmission electron microscopes. The improvements must concern
the brightness, the quantum yield, the workfunction and the chemical inertness and lifetime of the
photocathodes. Some of our recent works provide examples of related designs \cite{KNemeth10,JZTerdik12,ARuth13}.
In the first example, we have shown \cite{KNemeth10} using band structure calculations
that the angular distribution of emitted electrons can drastically
change when depositing an ultrathin (2-4 monolayers) MgO layer on the Ag(001) crystal surface. 
While a lot of electronic
structure studies have been carried out on the MgO:Ag(001) system to explain the experimentally observed 
variation of its catalytic
activity with varying number of MgO monolayers, we were the first to propose using this variation to
optimize the brightness (angular distribution) of emitted electrons through overlayer deposition.
Recent experimental results \cite{TCDroubay13} 
have confirmed our predictions and point out the drastic change in the
angular distribution of emitted electrons due to MgO overlayers on Ag(001).
Ultrathin oxide monolayers over metal surfaces change the electrostatic boundary conditions on the surface and thus they may
significantly decrease or increase the workfunctions and the shape and occupation of surface bands which leads to
drastic changes in the properties of the emitted electrons \cite{KNemeth10}.
In the second example, we have pointed out that the workfunction of the seasoned, high quantum efficiency 
Cs$_{2}$Te photoemissive material can be lowered from $\approx$ 3 eV to about 2.4 eV by acetylation
resulting in Cs$_{2}$TeC$_{2}$ while its quantum efficiency is preserved \cite{JZTerdik12}. 
Analogous compounds Cs$_{2}$PdC$_{2}$ and Cs$_{2}$PtC$_{2}$ exist and we
have managed to synthesize Li$_{2}$TeC$_{2}$ (x-ray diffraction confirms theoretically predicted
structure) \cite{Li2TeC2} 
as the first member of ternary acetylides with Te. This design was motivated by the goal of
turning Cs$_{2}$Te into an easier to excite $\pi$-electron system.
Our current research directions in the photocathode field focus on the theoretical screening of various inorganic
and organic overlayers on metals and semiconductors to lower workfunctions, optimize
brightness, quantum yield and life time of such
photocathodes in the often harsh temperature, imperfect vacuum and electromagnetic field environments they are used.

\section{Conclusions}
The present paper discusses several examples of materials design using quantum chemical and other theoretical /
computational methods to develop improved Li-ion and Li-air batteries and improved photoemissive
materials. The experimental testing of some of these designs is underway.

\section{Acknowledgements} 
Discussions with Drs. L. Shaw (IIT), K. C. Harkay and G. Srajer (Argonne) are gratefully acknowledged.
The Li-air battery and the photocathode research has been supported by the U.S. DOE Office of Science, under
contract No. DE-AC02-06CH11357 and NSF (No. PHY-0969989).

\footnotetext{\textit{$^{a}$~Address: Physics Department, Illinois Institute of Technology, 
Chicago, Illinois 60616, USA, nemeth@agni.phys.iit.edu}}

\footnotesize{

\begin{mcitethebibliography}{30}
\providecommand*{\natexlab}[1]{#1}
\providecommand*{\mciteSetBstSublistMode}[1]{}
\providecommand*{\mciteSetBstMaxWidthForm}[2]{}
\providecommand*{\mciteBstWouldAddEndPuncttrue}
  {\def\EndOfBibitem{\unskip.}}
\providecommand*{\mciteBstWouldAddEndPunctfalse}
  {\let\EndOfBibitem\relax}
\providecommand*{\mciteSetBstMidEndSepPunct}[3]{}
\providecommand*{\mciteSetBstSublistLabelBeginEnd}[3]{}
\providecommand*{\EndOfBibitem}{}
\mciteSetBstSublistMode{f}
\mciteSetBstMaxWidthForm{subitem}
{(\emph{\alph{mcitesubitemcount}})}
\mciteSetBstSublistLabelBeginEnd{\mcitemaxwidthsubitemform\space}
{\relax}{\relax}

\bibitem[Zalesny \emph{et~al.}(2011)Zalesny, Papadopoulos, Mezey, and
  Leszczynski]{LinearScaling}
\emph{Linear-Scaling Techniques in Computational Chemistry and Physics}, ed.
  R.~Zalesny, M.~G. Papadopoulos, P.~G. Mezey and J.~Leszczynski, Springer,
  Berlin, Heidelberg, New York, 1st edn, 2011, vol.~13\relax
\mciteBstWouldAddEndPuncttrue
\mciteSetBstMidEndSepPunct{\mcitedefaultmidpunct}
{\mcitedefaultendpunct}{\mcitedefaultseppunct}\relax
\EndOfBibitem
\bibitem[{P. Albertus et al.}(2010)]{BoschCO2patent10}
{P. Albertus et al.}, \emph{{High Specific Energy Li/O$_{2}$-CO$_{2}$ Battery;
  assignee: Bosch LLC; US Patent Application 12/907,205}}, 2010\relax
\mciteBstWouldAddEndPuncttrue
\mciteSetBstMidEndSepPunct{\mcitedefaultmidpunct}
{\mcitedefaultendpunct}{\mcitedefaultseppunct}\relax
\EndOfBibitem
\bibitem[{S. C. Jones et al.}(2011)]{CFx-ContourEnergy11}
{S. C. Jones et al.}, \emph{{Polymer Materials As Binder for a CFx Cathode;
  assignee: Contour Energy Systems Inc.; US Patent Application 13/010,431}},
  2011\relax
\mciteBstWouldAddEndPuncttrue
\mciteSetBstMidEndSepPunct{\mcitedefaultmidpunct}
{\mcitedefaultendpunct}{\mcitedefaultseppunct}\relax
\EndOfBibitem
\bibitem[Nemeth \emph{et~al.}(2013)Nemeth, van Veenendaal, and
  Srajer]{US8389178}
K.~Nemeth, M.~van Veenendaal and G.~Srajer, \emph{{Electrochemical energy
  storage device based on carbon dioxide as electroactive species, assignee:
  US. Dept. of Energy, Patent (granted), US8389178}}, 2013\relax
\mciteBstWouldAddEndPuncttrue
\mciteSetBstMidEndSepPunct{\mcitedefaultmidpunct}
{\mcitedefaultendpunct}{\mcitedefaultseppunct}\relax
\EndOfBibitem
\bibitem[Pearce(2012)]{JMPearce12}
J.~M. Pearce, \emph{Nature}, 2012, \textbf{491}, 519--521\relax
\mciteBstWouldAddEndPuncttrue
\mciteSetBstMidEndSepPunct{\mcitedefaultmidpunct}
{\mcitedefaultendpunct}{\mcitedefaultseppunct}\relax
\EndOfBibitem
\bibitem[{J. Hafner}(2008)]{VASP}
{J. Hafner}, \emph{J Comput Chem 29: 2044–2078, 2008}, 2008, \textbf{29},
  2044–2078\relax
\mciteBstWouldAddEndPuncttrue
\mciteSetBstMidEndSepPunct{\mcitedefaultmidpunct}
{\mcitedefaultendpunct}{\mcitedefaultseppunct}\relax
\EndOfBibitem
\bibitem[{P. Gianozzi {\it et.al}}(2009)]{QE}
{P. Gianozzi {\it et.al}}, \emph{J. Phys.: Condens. Matter}, 2009, \textbf{21},
  395502\relax
\mciteBstWouldAddEndPuncttrue
\mciteSetBstMidEndSepPunct{\mcitedefaultmidpunct}
{\mcitedefaultendpunct}{\mcitedefaultseppunct}\relax
\EndOfBibitem
\bibitem[{Paier et al.}(2006)]{JPaier06}
J.~{Paier et al.}, \emph{J. Chem. Phys.}, 2006, \textbf{124}, 154709\relax
\mciteBstWouldAddEndPuncttrue
\mciteSetBstMidEndSepPunct{\mcitedefaultmidpunct}
{\mcitedefaultendpunct}{\mcitedefaultseppunct}\relax
\EndOfBibitem
\bibitem[{Anisimov et al.}(1991)]{dftu}
V.~I. {Anisimov et al.}, \emph{Phys. Rev. B}, 1991, \textbf{44}, 943\relax
\mciteBstWouldAddEndPuncttrue
\mciteSetBstMidEndSepPunct{\mcitedefaultmidpunct}
{\mcitedefaultendpunct}{\mcitedefaultseppunct}\relax
\EndOfBibitem
\bibitem[{Soler, José M. et al.}(2002)]{Siesta}
{Soler, José M. et al.}, \emph{J. Phys. Cond. Mat.}, 2002, \textbf{14},
  2745\relax
\mciteBstWouldAddEndPuncttrue
\mciteSetBstMidEndSepPunct{\mcitedefaultmidpunct}
{\mcitedefaultendpunct}{\mcitedefaultseppunct}\relax
\EndOfBibitem
\bibitem[{{M. J. Frisch} et al.}(2013)]{Gaussian09}
{{M. J. Frisch} et al.}, \emph{{\sc Gaussian09}}, 2013, Gaussian, Inc.,
  Wallingford CT, http://www.gaussian.com\relax
\mciteBstWouldAddEndPuncttrue
\mciteSetBstMidEndSepPunct{\mcitedefaultmidpunct}
{\mcitedefaultendpunct}{\mcitedefaultseppunct}\relax
\EndOfBibitem
\bibitem[{Baker et al.}(2012)]{PQS}
J.~{Baker et al.}, \emph{WIREs Comput. Mol. Sci.}, 2012, \textbf{2}, 63\relax
\mciteBstWouldAddEndPuncttrue
\mciteSetBstMidEndSepPunct{\mcitedefaultmidpunct}
{\mcitedefaultendpunct}{\mcitedefaultseppunct}\relax
\EndOfBibitem
\bibitem[Bock \emph{et~al.}(2013)Bock, Challacombe, Gan, Henkelman, Nemeth,
  Niklasson, Odell, Schwegler, Tymczak, and Weber]{FreeON}
N.~Bock, M.~Challacombe, C.~K. Gan, G.~Henkelman, K.~Nemeth, A.~M.~N.
  Niklasson, A.~Odell, E.~Schwegler, C.~J. Tymczak and V.~Weber, \emph{{\sc
  FreeON}}, 2013, Los Alamos National Laboratory (LA-CC 01-2; LA-CC-04-086),
  Copyright University of California., http://www.freeon.org/\relax
\mciteBstWouldAddEndPuncttrue
\mciteSetBstMidEndSepPunct{\mcitedefaultmidpunct}
{\mcitedefaultendpunct}{\mcitedefaultseppunct}\relax
\EndOfBibitem
\bibitem[Suhai(1983)]{SSuhai83}
S.~Suhai, \emph{Phys. Rev. B}, 1983, \textbf{27}, 3506\relax
\mciteBstWouldAddEndPuncttrue
\mciteSetBstMidEndSepPunct{\mcitedefaultmidpunct}
{\mcitedefaultendpunct}{\mcitedefaultseppunct}\relax
\EndOfBibitem
\bibitem[N{\'e}meth and Srajer(2014)]{KNemeth13}
K.~N{\'e}meth and G.~Srajer, \emph{RSC Advances}, 2014, \textbf{4}, 1879\relax
\mciteBstWouldAddEndPuncttrue
\mciteSetBstMidEndSepPunct{\mcitedefaultmidpunct}
{\mcitedefaultendpunct}{\mcitedefaultseppunct}\relax
\EndOfBibitem
\bibitem[Melot and Tarascon(2013)]{BCMelot13}
B.~C. Melot and J.-M. Tarascon, \emph{Acc. Chem. Res.}, 2013, \textbf{46},
  1226--1238\relax
\mciteBstWouldAddEndPuncttrue
\mciteSetBstMidEndSepPunct{\mcitedefaultmidpunct}
{\mcitedefaultendpunct}{\mcitedefaultseppunct}\relax
\EndOfBibitem
\bibitem[{Hautier et al.}(2011)]{GHautier11}
G.~{Hautier et al.}, \emph{J. Mater. Chem.}, 2011, \textbf{21},
  17147–17153\relax
\mciteBstWouldAddEndPuncttrue
\mciteSetBstMidEndSepPunct{\mcitedefaultmidpunct}
{\mcitedefaultendpunct}{\mcitedefaultseppunct}\relax
\EndOfBibitem
\bibitem[{Zhou et al.}(2004)]{FZhou04}
F.~{Zhou et al.}, \emph{Electrochem. Comm.}, 2004, \textbf{6}, 1144--1148\relax
\mciteBstWouldAddEndPuncttrue
\mciteSetBstMidEndSepPunct{\mcitedefaultmidpunct}
{\mcitedefaultendpunct}{\mcitedefaultseppunct}\relax
\EndOfBibitem
\bibitem[{Jain et al.}(2011)]{AJain11}
A.~{Jain et al.}, \emph{Phys. Rev. B}, 2011, \textbf{84}, 045115\relax
\mciteBstWouldAddEndPuncttrue
\mciteSetBstMidEndSepPunct{\mcitedefaultmidpunct}
{\mcitedefaultendpunct}{\mcitedefaultseppunct}\relax
\EndOfBibitem
\bibitem[Nemeth(2013)]{BNpatent}
K.~Nemeth, \emph{{Functionalized Boron Nitride Materials as Electroactive
  Species in Electrochemical Energy Storage Devices, patent pending, assignee:
  Nemeth's Materials Design LLC}}, 2013\relax
\mciteBstWouldAddEndPuncttrue
\mciteSetBstMidEndSepPunct{\mcitedefaultmidpunct}
{\mcitedefaultendpunct}{\mcitedefaultseppunct}\relax
\EndOfBibitem
\bibitem[{Yamane et al.}(1987)]{HYamane87}
H.~{Yamane et al.}, \emph{J. Solid State Chem.}, 1987, \textbf{71}, 1--11\relax
\mciteBstWouldAddEndPuncttrue
\mciteSetBstMidEndSepPunct{\mcitedefaultmidpunct}
{\mcitedefaultendpunct}{\mcitedefaultseppunct}\relax
\EndOfBibitem
\bibitem[Perdew \emph{et~al.}(1996)Perdew, Burke, and Ernzerhof]{PBE}
J.~P. Perdew, K.~Burke and M.~Ernzerhof, \emph{Phys. Rev. Lett.}, 1996,
  \textbf{77}, 3865\relax
\mciteBstWouldAddEndPuncttrue
\mciteSetBstMidEndSepPunct{\mcitedefaultmidpunct}
{\mcitedefaultendpunct}{\mcitedefaultseppunct}\relax
\EndOfBibitem
\bibitem[{Perdew et al.}(2008)]{PBEsol}
J.~P. {Perdew et al.}, \emph{Phys. Rev. Lett.}, 2008, \textbf{100},
  136406\relax
\mciteBstWouldAddEndPuncttrue
\mciteSetBstMidEndSepPunct{\mcitedefaultmidpunct}
{\mcitedefaultendpunct}{\mcitedefaultseppunct}\relax
\EndOfBibitem
\bibitem[{Mason et al.}(2011)]{THMason11}
T.~H. {Mason et al.}, \emph{J. Phys. Chem. C}, 2011, \textbf{115},
  16681--16688\relax
\mciteBstWouldAddEndPuncttrue
\mciteSetBstMidEndSepPunct{\mcitedefaultmidpunct}
{\mcitedefaultendpunct}{\mcitedefaultseppunct}\relax
\EndOfBibitem
\bibitem[{Huang et al.}(2008)]{JHuang08}
J.~{Huang et al.}, \emph{Phys. Chem. Chem. Phys.}, 2008, \textbf{10},
  2625\relax
\mciteBstWouldAddEndPuncttrue
\mciteSetBstMidEndSepPunct{\mcitedefaultmidpunct}
{\mcitedefaultendpunct}{\mcitedefaultseppunct}\relax
\EndOfBibitem
\bibitem[{{K. N{\'e}meth, K.C. Harkay \it et.al}}(2010)]{KNemeth10}
{{K. N{\'e}meth, K.C. Harkay \it et.al}}, \emph{Phys. Rev. Lett.}, 2010,
  \textbf{104}, 046801\relax
\mciteBstWouldAddEndPuncttrue
\mciteSetBstMidEndSepPunct{\mcitedefaultmidpunct}
{\mcitedefaultendpunct}{\mcitedefaultseppunct}\relax
\EndOfBibitem
\bibitem[Terdik and {N{\'e}meth et al.}(2012)]{JZTerdik12}
J.~Z. Terdik and K.~{N{\'e}meth et al.}, \emph{Phys. Rev. B}, 2012,
  \textbf{86}, 035142\relax
\mciteBstWouldAddEndPuncttrue
\mciteSetBstMidEndSepPunct{\mcitedefaultmidpunct}
{\mcitedefaultendpunct}{\mcitedefaultseppunct}\relax
\EndOfBibitem
\bibitem[Ruth and {N{\'e}meth et al.}(2013)]{ARuth13}
A.~Ruth and K.~{N{\'e}meth et al.}, \emph{J. Appl. Phys.}, 2013, \textbf{113},
  183703\relax
\mciteBstWouldAddEndPuncttrue
\mciteSetBstMidEndSepPunct{\mcitedefaultmidpunct}
{\mcitedefaultendpunct}{\mcitedefaultseppunct}\relax
\EndOfBibitem
\bibitem[{Droubay et al.}(2013)]{TCDroubay13}
T.~C. {Droubay et al.}, \emph{Phys. Rev. Lett., submitted}, 2013\relax
\mciteBstWouldAddEndPuncttrue
\mciteSetBstMidEndSepPunct{\mcitedefaultmidpunct}
{\mcitedefaultendpunct}{\mcitedefaultseppunct}\relax
\EndOfBibitem
\bibitem[N{\'e}meth \emph{et~al.}()N{\'e}meth, Unni, and {J. Kaduk et
  al.}]{Li2TeC2}
K.~N{\'e}meth, A.~K. Unni and {J. Kaduk et al.}, \emph{{The synthesis of
  ternary acetylides with Te, to be published}}\relax
\mciteBstWouldAddEndPuncttrue
\mciteSetBstMidEndSepPunct{\mcitedefaultmidpunct}
{\mcitedefaultendpunct}{\mcitedefaultseppunct}\relax
\EndOfBibitem
\end{mcitethebibliography}
\bibliographystyle{rsc}
\providecommand*{\mcitethebibliography}{\thebibliography}
\csname @ifundefined\endcsname{endmcitethebibliography}
{\let\endmcitethebibliography\endthebibliography}{}

}
\end{document}